\documentclass[prb,preprint]{revtex4-1} 
\usepackage{amsmath} 
\usepackage{amsfonts}
\usepackage{graphicx}
\usepackage{hyperref}
\usepackage{ifthen}

\usepackage{tabularx}
    \newcolumntype{L}{>{\raggedright\arraybackslash}X}
\usepackage{color, colortbl}
\definecolor{Gray}{gray}{0.9}
\usepackage{setspace}
\usepackage[colorinlistoftodos]{todonotes}

\newboolean{anon_flag}
\setboolean{anon_flag}{False} 

\begin{document}

\title{Data Science Education in Undergraduate Physics: Lessons Learned from a Community of Practice}
\ifthenelse{\boolean{anon_flag}}{
  % Anonymous version
}{
\author{Karan Shah}
\email{k.shah@hzdr.de}
\affiliation{Center for Advanced Systems Understanding, Helmholtz-Zentrum Dresden-Rossendorf, G\"orlitz, Germany}

\author{Julie Butler}
\email{butlerju@mountunion.edu}
\affiliation{Department of Biochemistry, Chemistry, and Physics, University of Mount Union, Alliance, Ohio 44601, USA}

\author{Alexis Knaub}
\email{avknaub@gmail.com}
\affiliation{American Association of Physics Teachers}

\author{Anıl Zenginoğlu}
\email{anil@umd.edu}
\affiliation{Institute for Physical Science and Technology, University of Maryland, College Park, Maryland 20742, USA}

\author{William Ratcliff}
\email{william.ratcliff@nist.gov}
\affiliation{National Institute of Standards and Technology}
\affiliation{Department of Physics, University of Maryland, College Park, Maryland 20742, USA}
\affiliation{Department of Materials Science and Engineering, University of Maryland, College Park, Maryland 20742, USA}
\author{Mohammad Soltanieh-ha}
\email{msoltani@bu.edu}
\affiliation{Questrom School of Business at Boston University}
}

%\date{\today}

\begin{abstract}

It is becoming increasingly important that physics educators equip their students with the skills to work with data effectively. However, many educators may lack the necessary training and expertise in data science to teach these skills. To address this gap, we created the Data Science Education Community of Practice (DSECOP), bringing together graduate students and physics educators from different institutions and backgrounds to share best practices and lessons learned from integrating data science into undergraduate physics education. In this article we present insights and experiences from this community of practice, highlighting key strategies and challenges in incorporating data science into the introductory physics curriculum. Our goal is to provide guidance and inspiration to educators who seek to integrate data science into their teaching, helping to prepare the next generation of physicists for a data-driven world.

\end{abstract}
\maketitle
\section{Introduction}

\begin{figure}[h]
    \centering
    \includegraphics[scale=0.5]{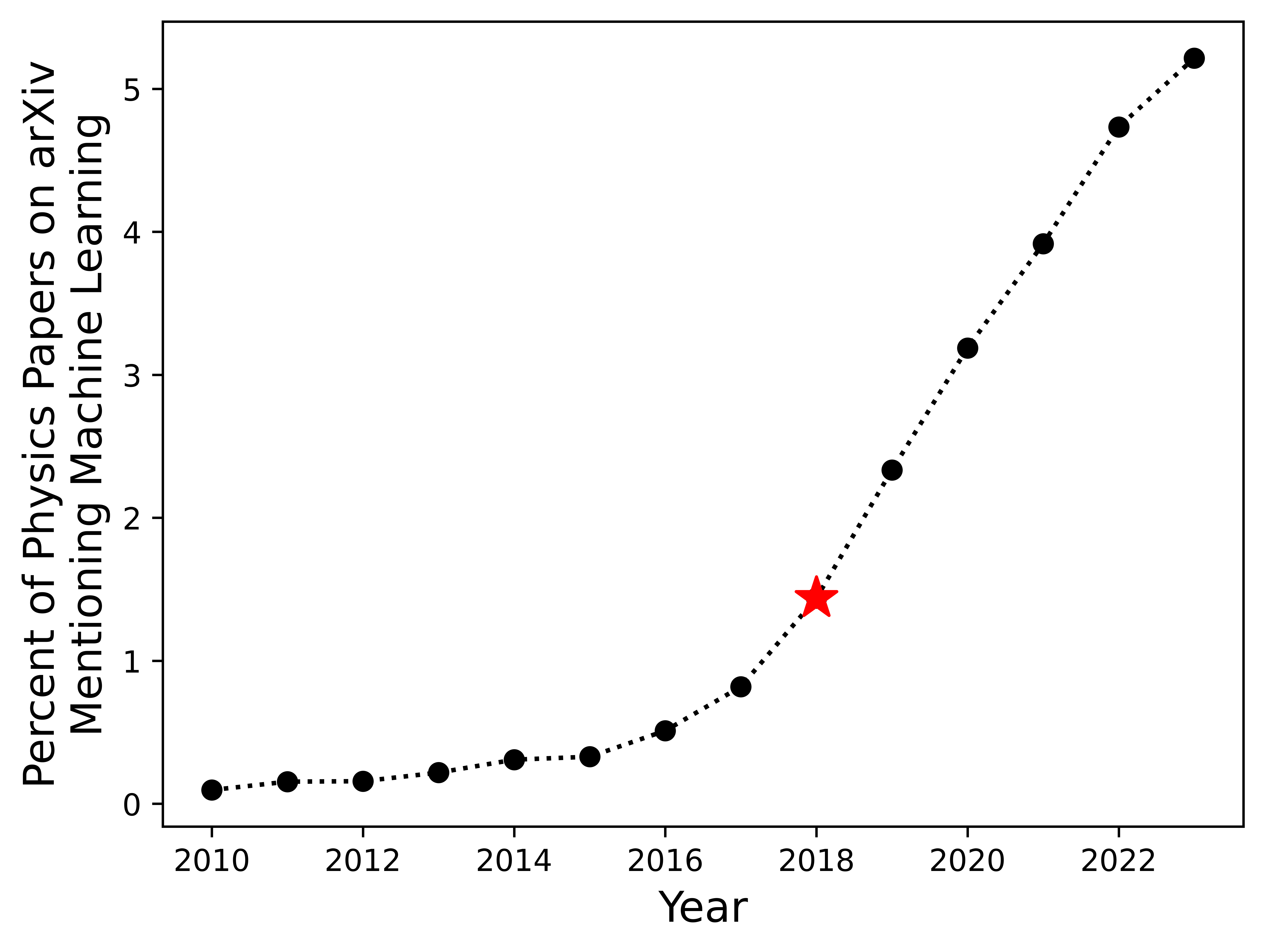}
    \caption{Percentage of academic papers per year uploaded to arXiv's physics section that mention ``machine learning" anywhere in the article. The red star represents the inception of the American Physical Society's Topical Group on Data Science (GDS) in 2018.}
    \label{fig:papers_by_year}
\end{figure}

Data analysis has always been an essential component of the scientific method. From the derivation of Kepler's laws of planetary motion by analysis of Brahe's astronomical observations to the detection of the Higgs boson, data analysis has played a crucial role in advancing our understanding of the natural world. Traditionally, data analysis in physics primarily focused on applying theoretical concepts to structured datasets derived from experiments or simulations. In the current age of big data, physicists have access to an unprecedented volume and variety of data, enabling new discoveries that would have been impossible without advanced data analysis techniques. These discoveries require not just analytical skills but also proficiency in advanced data management, statistical modeling, and computational techniques, skills that are not traditionally taught in undergraduate physics classes. The historical focus on theoretical and experimental physics leaves little room for data science education with physics applications. Additionally, many physics educators lack the training and expertise to teach data science effectively, often because of the fast-paced evolution of tools and techniques. As illustrated in Figure \ref{fig:papers_by_year}, the frequency of discussion of data science in physics research and industry has been growing rapidly. Integrating data science into undergraduate physics curricula will prepare students for the changing demands of the future.

It is difficult to define data science comprehensively,  partly because many definitions tend to emphasize the specific activities of practitioners within their respective fields \cite{provost2013data}.  However, we will provide an operational definition. The term was first used in the 1960s in military and corporate contexts \cite{alvarado_data_2023}. While data science is an umbrella term \cite{Irizarry2020Role}, the NIST Big Data Interoperability Framework \cite{nist_big_data_public_working_group_definitions_and_taxonomies_subgroup_nist_2015} provides a succinct definition covering various aspects of data science:

``Data science is the extraction of actionable knowledge directly from data through a process of discovery, or hypothesis formulation and hypothesis testing.''

Data science uses statistical and computational methods to extract insights from data. It includes data collection, cleaning, preparation, exploratory data analysis, statistical modeling and inference, machine learning (ML), and data visualization. It has been used to address questions in various domains, such as biology, education, physics, business, linguistics, and medicine \cite{oliver2021undergraduate}. Machine Learning (ML) is a subset of data science that involves the development of algorithms and statistical models that enable computers to automatically extract patterns from data. \cite{Deisenroth2020, Goodfellow-et-al-2016}.

Data science analyzes large and complex datasets to identify patterns and trends, gain insights, and make probabilistic predictions. Machine learning models already serve as faster and more generalizable surrogate models, replacing theory in a variety of domains \cite{carleoMachineLearningPhysical2019a, karniadakisPhysicsinformedMachineLearning2021b}. However, a key limitation of many current ML models is their lack of interpretability, often functioning as `black boxes' that are not readily explainable. In the future, ML models may represent reality with interpretable models as theoretical physics does\cite{rudin_interpretable_2022}. While we are far from that future, it is an exciting research direction for physicists.

% It is the right time to introduce data science in undergraduate physics
\section{Motivation}

In this section, we argue that data science should be part of the undergraduate physics curriculum. Data science is useful not only for physics research but also for careers in industry and as a pedagogical tool to improve physics education.

\subsection{Why is data science important, particularly for physics undergraduates?} 

Physics research is increasingly data-intensive (Data Science for Physics): Many fields of physics, such as astronomy, particle physics, condensed matter physics, and biophysics, produce vast amounts of data that require sophisticated data analysis. In addition to theory, experiment, and simulations, data science is the ``fourth pillar" of science and has a place in the scientific method \cite{hey2009the}. With the rise of massive data-intensive experiments and simulations, it has become essential that physicists employ statistical and machine learning techniques in their workflow. Data science proficiency is needed to write programs for sifting through and generating insights from experiments and large-scale simulations that generate terabytes of data.  Additionally, these skills can enhance the efficiency of experiments and simulations, allowing for fewer measurements while still answering physically meaningful questions.  Some domains where data science is used are collider physics \cite{hajer2020novelty}, quantum physics \cite{das2019machine},  the search for new materials, and the control of fusion reactors \cite{degraveMagneticControlTokamak2022}.

The increasing importance of data science in the workforce (Data Science for Future Careers): In the era of Industry 4.0\cite{schwabFourthIndustrialRevolution2017}, where automation, data exchange, and interconnectedness shape our global industries, data science skills have become paramount. Physics undergraduates equipped with these skills are well-positioned to navigate the challenges and opportunities of this new industrial landscape. It is essential to note that incorporating data science doesn't detract from core physics topics. Instead, it provides an enriched framework to understand and apply these principles using contemporary techniques. As industries pivot towards data-driven decision-making, grounding physics students in data science amplifies their analytical prowess and broadens their career horizons. 

Data science can enhance physics education (Data Science for Education): Due to curriculum pressure, it is difficult to target parts of physics courses to replace data science education. Introducing data science should not take away from the fundamental physics curriculum but, on the contrary, should make it easier to teach and demonstrate the fundamental ideas using modern data science tools. Teaching data science to physics students should not simply replace the existing physics curriculum with new content but try to enhance it using data science as a tool. Data science offers new and innovative ways to teach physics concepts and engage students in active learning by providing hands-on experience with physics-relevant data sets.
%Free and open-source software, online tutorials, and educational materials make it easier than ever to introduce data science in undergraduate physics education.

Growing overlap between physical concepts and machine learning techniques (Physics for Data Science): Recent research has demonstrated that physical ideas, such as diffusion, symmetry, or relativistic geometry, can be used to develop more powerful and efficient machine learning algorithms and data analysis techniques \cite{nickel2018learning, carleoMachineLearningPhysical2019a, bahri_statistical_2020}. By exploring the connections between physics and data science, students can gain insights into the fundamental principles that underlie both fields and develop a more holistic view of the role of data in scientific discovery and innovation. The relationship between data science and physics is bidirectional. While data science provides powerful tools for analyzing complex physical phenomena, physics, in turn, provides a rich context for developing and refining data science methodologies \cite{carleoMachineLearningPhysical2019a}. Physics-informed machine learning is a rapidly developing field combining black-box machine learning models with physics constraints \cite{karniadakisPhysicsinformedMachineLearning2021b}.

\subsection{Why is it important for physics professors to teach data science (as opposed to computer scientists)?}

Existing undergraduate data science programs focus primarily on theoretical foundations and quantitative skills with very little domain knowledge outside of computer science \cite{oliver2021undergraduate}. Graduates of these programs may lack the appropriate context for designing and evaluating domain-specific data science applications. The  National Academies of Sciences, Engineering \& Medicine (NASEM) framework emphasizes the importance of domain knowledge for effectively applying data science \cite{NAP25104}. Physics professors can provide the substantial domain knowledge needed to make data science accessible to physics students \cite{Berthold2019What, Irizarry2020Role}.

\subsection{Survey Results Suggest Data Science is typically not part of the undergraduate physics curriculum}

In March 2022, we ran two surveys to better understand the data science and undergraduate physics education landscape. We first surveyed faculty to learn whether they were teaching data science in their undergraduate physics courses and, if so, in which courses and what data science skills. Next, we surveyed data scientists in the industry who are involved with hiring physics bachelors to understand what data science skills and knowledge they would expect a recent bachelor's in physics to have for data science jobs. The surveys were based on the \ifthenelse{\boolean{anon_flag}}{COP}{DSECOP} leadership's experience in data science and physics, as well as a previous local survey \cite{losert2020}. Each survey was pilot-tested to ensure coherence and inclusion of important questions. The surveys were sent through American Physical Society (APS) listservs (e.g., Forum on Education [FEd], GDS), message boards, and our various professional networks. 

\textit{Faculty Survey}

We had 100 physics faculty responses from 78 US institutions (5 US institutions had two respondents), 7 international institutions, and 10 unknown  \footnote{We do not have a response rate as the surveys were sent through American Physical Society (APS) listservs (e.g., Forum on Education [FEd], GDS), message boards, and our various professional networks. We assume the response rate was not high, given that the APS GDS list-serv alone has over 2000 members. Despite the low response rate, we gained valuable information from the completed surveys.}. Of the 30 respondents who said they had taught data science in undergraduate physics courses in the past five years, 25 are from US institutions and 5 are from international institutions. Those who had taught data science reported teaching it in introductory, intermediate, and advanced courses, and in general courses and those focused on data science. Because more instructors reported teaching it in non-introductory courses, we will focus on those first. For such courses, data science is taught via in-class activities (\textit{N} = 8), projects (\textit{N} = 25), and homework exercises (\textit{N} = 19). Some respondents include data science questions on quizzes or exams (\textit{N} = 8).

%For introductory courses, data science is taught via in-class activities (\textit{N} = 20), projects (\textit{N} = 8), and homework exercises (\textit{N} = 7). A few respondents include data science on quizzes or exams (\textit{N} = 3). 

\begin{figure}
\centering
\includegraphics[width=\linewidth]{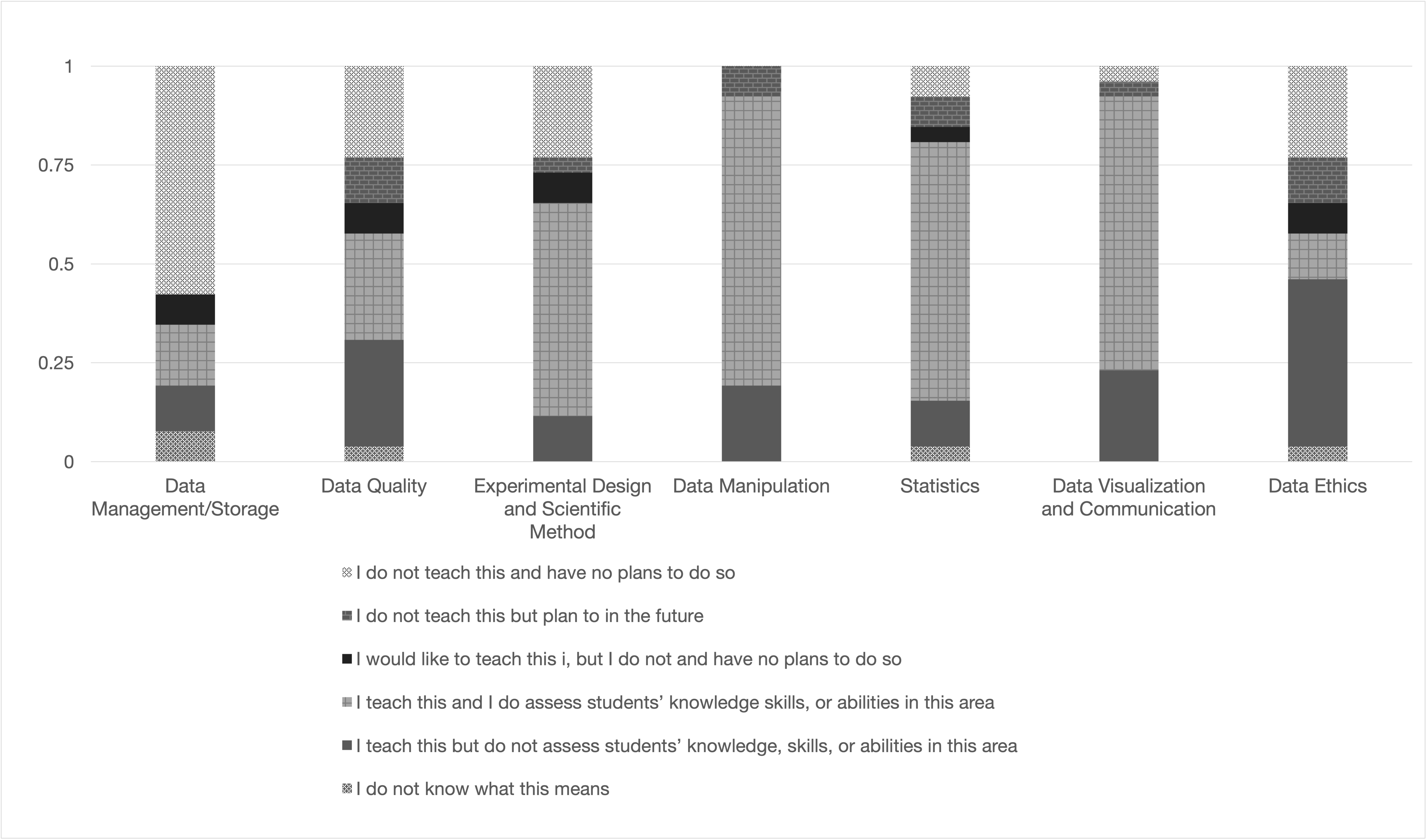}
\caption{Results of data skills survey for faculty who teach data science in their intermediate/advanced physics courses. Relevant skills are listed on the x-axis, with the different textures showing the proportion of responses received.}
\label{fig:SkillsFaculty}
\end{figure}

Figure \ref{fig:SkillsFaculty} is a plot of the data skills and the statement that the respondents selected (\textit{N} = 26). We find that more faculty teach and assess experimental design, data collection, data manipulation, statistics, and data visualization, and fewer teach data management and storage.

\textit{Industry Survey}

Twenty-five individuals from 22 different employers responded to this survey. The respondents worked at companies of varying sizes, ranging from one employee to thousands. Most worked in government scientific research (\textit{N} = 10) or in tech (\textit{N} = 7). Others worked in for-profit research (general), finance, advertising, health care, and social media, with most working in companies that covered multiple areas. Most (\textit{N} = 19 or 76\%) worked with data scientists with physics degrees. 

\begin{figure}
\centering
\includegraphics[width=\linewidth]{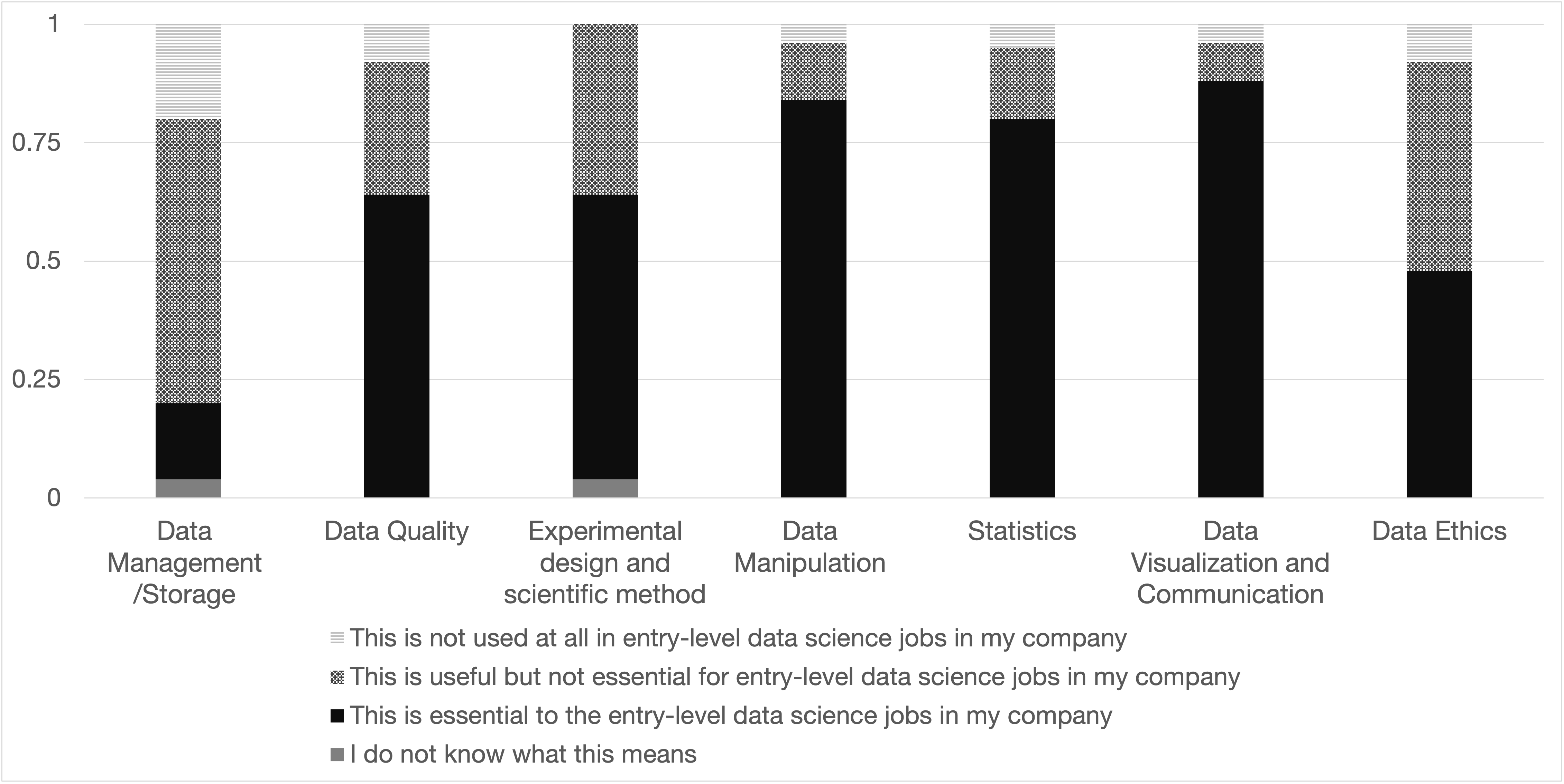}
\caption{Results of data skills survey for industry practitioners. Relevant skills are listed on the x-axis, with the different textures showing the proportion of responses received.}
\label{fig:IndustryPlot}
\end{figure}

Respondents were asked about various technical skills and knowledge they would expect from an entry-level data scientist with a bachelor's in physics. Most (\textit{N} = 20 or 80\%) would expect such an employee to be programming daily, with slightly fewer (\textit{N} = 18 or 72\%) using various data science software packages (e.g., NumPy \cite{harris2020array}, SciPy \cite{2020SciPy-NMeth}, Pandas \cite{reback2020pandas}) daily. Almost all respondents indicated that such employees use Python (\textit{N} = 22, 88\%), though many (\textit{N} = 16, 64\%)  use C/C++ and Matlab (\textit{N} = 15, 60\%). As shown in Figure \ref{fig:IndustryPlot}, over half of the industry respondents indicated that the following were essential for an entry-level data scientist job: Data Manipulation  (e.g., transforming data, gaining insights from data); Experimental design and scientific method (e.g., designing experiments, data collection); Statistics (e.g., model selection and uncertainty, A/B testing); and Data Visualization and Communication (e.g., plotting data, technical writing).

Additionally, some respondents left further clarifying comments. They emphasized that tools (e.g., programming languages) are less important than understanding how to make useful plots or understanding the bigger goals. Good collaboration and communication skills and best practices for software development and error analysis were also mentioned.

\section{Challenges}

%% Not all professors have the necessary background
The survey of instructors revealed several challenges in incorporating data science into the undergraduate physics curriculum. First, many physicists who teach undergraduate courses are not familiar enough with data science topics to teach them, and preparing such new content is extremely time-consuming for them.

%% If you add one thing, you have to cut another thing
A second challenge is finding a place for data science in the normal course structure. Especially for courses taught in sequence over multiple semesters, a well-established flow of topics leaves little room for new material to be added. 

A third challenge is that many universities lack the resources to offer a new class specifically on data science topics, and many students lack the flexibility in their schedules to take an additional course. This lack of resources is particularly prohibitive in developing countries, where financial constraints are more limiting. Universities often struggle with the high cost of hardware infrastructure and proprietary software, which may seem essential for teaching data science.
   
%% Knowledge of students
The final challenge we wish to discuss is the required background students must have outside the physics curriculum. To fully understand much of machine learning, a higher level of mathematics, statistics, and programming is needed. Some of these skills may be picked up in mathematics or computer science courses but are not typically taught in the physics classroom. While programming knowledge is becoming more common among physics students (even becoming a requirement in some departments), this is not necessarily true for all universities and thus hinders incorporating data science into the physics classroom when students do not have the necessary background knowledge.

\ifthenelse{\boolean{anon_flag}}{\section{The Community of Practice (COP)}}{\section{Data Science Education Community of Practice (DSECOP)}}
%% Mission Statement

Link: \href{https://dsecop.org}{https://dsecop.org}

The mission of \ifthenelse{\boolean{anon_flag}}{COP}{DSECOP} \ifthenelse{\boolean{anon_flag}}{}{\cite{dsecop_web}} is to support physics educators who wish to integrate data science into their existing courses by providing resources as well as a community of like-minded educators for support. \ifthenelse{\boolean{anon_flag}}{The COP project came to life in 2022 following the awarding of a prestigious grant.}{The DSECOP project came to life in 2022 following the awarding of the competitive Innovation Fund Award from the American Physical Society to some leaders of the Topical Group of Data Science (GDS) \cite{aps_gds}.}

%% Modules Purpose
One of the resources provided by the \ifthenelse{\boolean{anon_flag}}{COP}{DSECOP} organization is a collection of free and open-source modules created by the group's fellows who are Ph.D. students and recent Ph.D. graduates. The modules do not require access to large datasets, run on free software, and can be executed from a browser. Each module introduces a specific topic in the realm of data science applied to a physics concept taught in the traditional physics major. The goal is that these modules are detailed enough to provide instructors and students with a good understanding of how data science can be used to solve a problem in physics but also short enough to be incorporated into a pre-existing course without significant alterations. These modules consist of a lesson plan, exercises, and solutions for the instructor.  We adopted this approach because not all universities will have the resources to develop a full course on data science.  This also allows for the content to be introduced gradually into the curriculum by individual professors rather than requiring approval for a new course that has to replace a well-established physics course.  It also presents alternative materials to the students for self-study that build on the parts that are taught in the classroom.  

Our fellows present an initial idea for a module and, after refining it with discussion, send it out for outside feedback from faculty.  They then develop a unit, which is later sent out for testing by other faculty.   We also run an annual workshop where the presentation of work by fellows and faculty is followed by a discussion. We run a webinar series to present different approaches to teaching data science, with topics ranging from ethics to careers in industry.

%% Overview of DSECOP and the Existing Modules

%% What level of physics and data science are the modules (introduce the DSECOP Numbering System)
\ifthenelse{\boolean{anon_flag}}{\subsection{COP Modules}}{\subsection*{DSECOP Modules}}

% \TODO{Can we provide the link more easily--maybe as a reference?}

\ifthenelse{\boolean{anon_flag}}{GitHub Link: (hidden for the anonymous version)}{Link: \url{https://github.com/GDS-Education-Community-of-Practice/DSECOP}\cite{dsecop_repo}}
\ifthenelse{\boolean{anon_flag}}{
\begin{figure}
    \centering
    \includegraphics[scale=0.36]{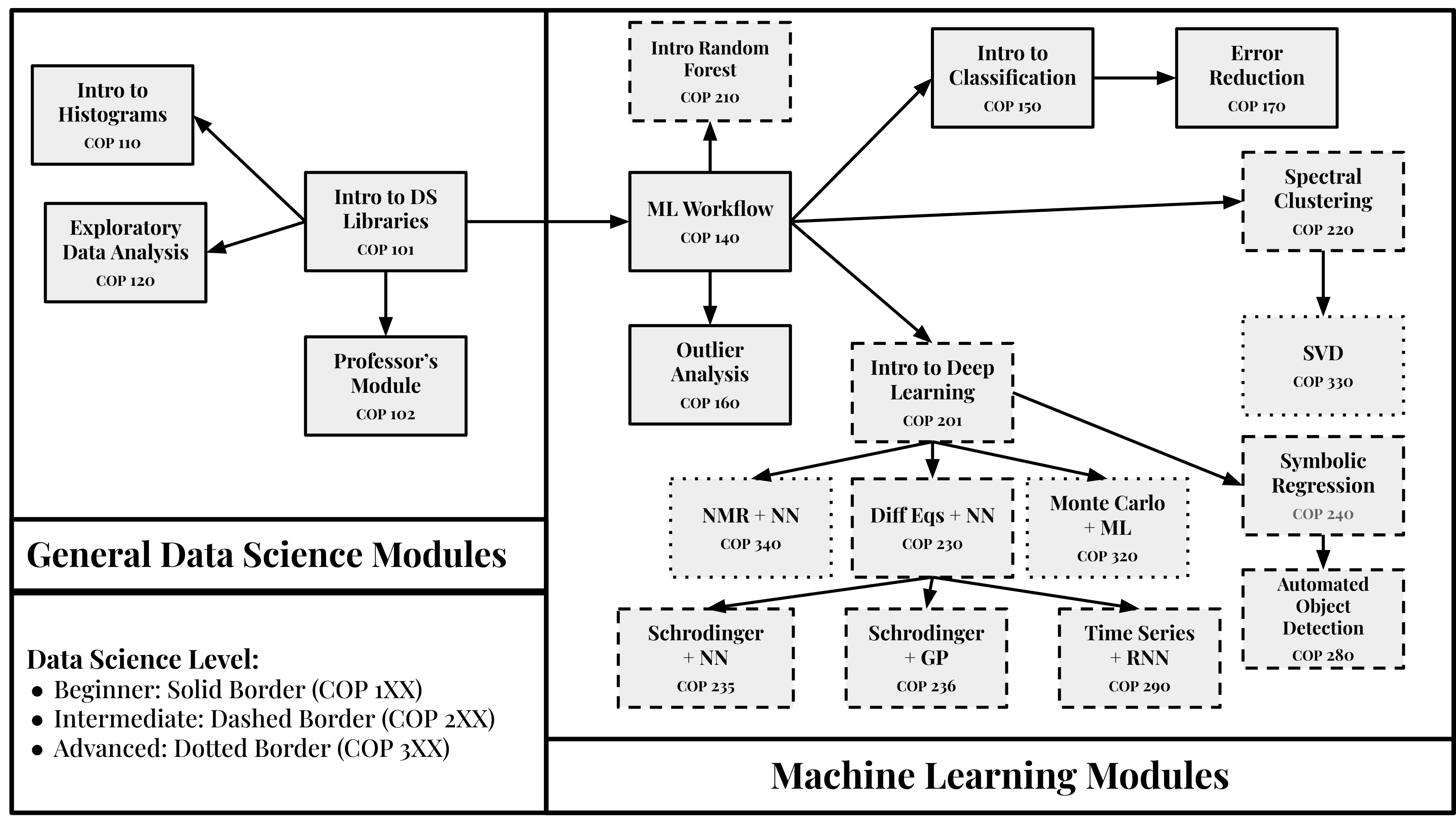}
    \caption{This flowchart shows the relationships between the modules and their topic (general data science or machine learning). The arrows depict suggested prerequisites for the relevant data science and machine learning topics, starting from the most basic module COP 101: Introduction to Data Science Libraries. Modules that cover data science and machine learning at the introductory level are shown with solid borders and are numbered COP 1XX. Modules with intermediate topics have dashed borders and are numbered COP 2XX and modules that cover advanced topics in data science and machine learning have dotted borders and are numbered COP 3XX.}
    \label{fig:dsecop_flowchart}
\end{figure}
}{
\begin{figure}
    \centering
    \includegraphics[scale=0.3]{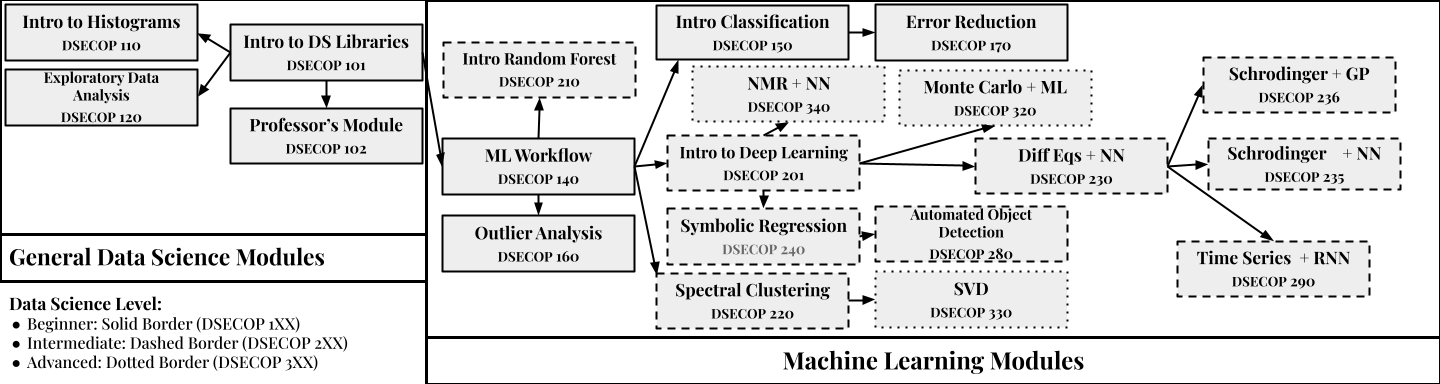}
    \caption{This flowchart shows the relationships between the modules and their topic (general data science or machine learning). The arrows depict suggested prerequisites for the relevant data science and machine learning topics, starting from the most basic module DSECOP 101: Introduction to Data Science Libraries. Introductory, intermediate, and advanced-level modules are denoted by their borders (solid, dashed, and dotted) and also by their module number (1XX, 2XX, 3XX).}
    \label{fig:dsecop_flowchart}
\end{figure}
}

The \ifthenelse{\boolean{anon_flag}}{COP}{DSECOP} modules span a wide range of data science concepts; thus, we have sought to order them regarding the level of data science knowledge. We introduce the \ifthenelse{\boolean{anon_flag}}{COP}{DSECOP} numbering system, where a 100-level module is appropriate for new and beginner programmers, a 200-level module includes intermediate-level machine learning concepts with 100-level prerequisites, and a 300-level module introduces advanced machine learning concepts. A roadmap of the \ifthenelse{\boolean{anon_flag}}{COP}{DSECOP} modules and their suggested prerequisite modules is shown in Figure \ref{fig:dsecop_flowchart}. This flowchart breaks the modules up into those that contain general data science topics and those that contain machine learning topics. 

To streamline the usage of modules, we establish workflows and conventions for both module creators and faculty members. We assume basic Python knowledge and standardize the list of libraries such as Matplotlib, NumPy, SciPy, Pandas, and TensorFlow. Setting up Python environments across different machines can be challenging. Additionally, not all students have access to GPUs, which are integral to modern data science tasks. To address these issues, all the modules can be run on the Google Colab platform \cite{GoogleColaboratory}, allowing students to run the modules in their browsers on a cloud GPU without any extra setup. The module can also be run in a local environment with the Conda framework for easy setup. Detailed instructions are provided for both the cloud-based Colab and local Conda methods.

At the time of this paper's publication, we have 21 posted modules spanning many common courses in the undergraduate physics curriculum. Table \ref{tab:dsecop_courses} breaks down the current modules by their physics content and divides them among the common courses in a physics curriculum. Note that the modules ``\ifthenelse{\boolean{anon_flag}}{COP}{DSECOP} 201: Introduction to Deep Learning" and ``\ifthenelse{\boolean{anon_flag}}{COP}{DSECOP} 210: Introduction to Random Forests" do not contain any physics content and thus may be introduced in any physics course with the proper computational background. 

\ifthenelse{\boolean{anon_flag}}
{
\begin{spacing}{0.8}
\begin{table}
    \centering
    \begin{tabularx}{\linewidth}{|L|L|L|} \hline
        \rowcolor{Gray}
         \textbf{General Physics (Lecture and Lab} & \textbf{Waves and Optics} & \textbf{Classical Mechanics} \\ \hline
         \textbf{COP 101}: Intro to DS Libraries \newline \textbf{COP 110}: Intro to Histograms \newline textbf{COP 120}: Exploratory Data Analysis \newline \textbf{COP 140}: ML Workflow  &  \textbf{COP 150}: Intro to Classification \newline \textbf{COP 160}: Outlier Analysis \newline
         \textbf{COP 170}: Error Reduction & \textbf{COP 220}: Spectral Clustering \newline \textbf{COP 230}: Diff Eqs + NN \newline \textbf{COP 290}: Time Series + RNN \\ \hline
        \rowcolor{Gray}
          \textbf{Electricity and Magnetism} & \textbf{Quantum Mechanics} & \textbf{Statistical Thermodynamics} \\ \hline 
         \textbf{COP 190}: Data Science with Griffiths \newline \textbf{COP 250}: Automatic Differentiation& \textbf{COP 235}: Schrodinger + NN \newline \textbf{COP 236}: Schrodinger + GP & \textbf{COP 320}: Monte Carlo + NN \\ \hline
        \rowcolor{Gray}
         \textbf{Advanced Lab} & \textbf{Particle, Nuclear, or Medical Physics} & \textbf{ Any Course} \\ \hline
         \textbf{COP 120}: Exploratory Data Analysis \newline \textbf{COP 240}: Symbolic Regression \newline \textbf{COP 280}: Automated Video Analysis \newline \textbf{COP 330}: SVD & \textbf{COP 101}: Intro to DS Libraries \newline \textbf{COP 110}: Intro to Histograms \newline\textbf{COP 340}: NMR + NN & \textbf{COP 102}: The Professor’s Module \newline \textbf{COP 201}: Intro to Deep Learning \newline \textbf{COP 210}: Intro to Random Forests \\ \hline
    \end{tabularx}
    \caption{An overview of courses in the undergraduate physics curriculum and existing modules for each course.}
    \label{tab:dsecop_courses}
\end{table}
\end{spacing}
}{
\begin{spacing}{0.8}
\begin{table}
    \centering
    \begin{tabularx}{\linewidth}{|L|L|L|} \hline
        \rowcolor{Gray}
         \textbf{General Physics (Lecture and Lab} & \textbf{Waves and Optics} & \textbf{Classical Mechanics} \\ \hline
         \textbf{DSECOP 101}: Intro to DS Libraries \newline \textbf{DSECOP 110}: Intro to Histograms \newline \textbf{DSECOP 120}: Exploratory Data Analysis \newline \textbf{DSECOP 140}: ML Workflow  &  \textbf{DSECOP 150}: Intro to Classification \newline \textbf{DSECOP 160}: Outlier Analysis \newline
         \textbf{DSECOP 170}: Error Reduction & \textbf{DSECOP 220}: Spectral Clustering \newline \textbf{DSECOP 230}: Diff Eqs + NN \newline \textbf{DSECOP 290}: Time Series + RNN \\ \hline
        \rowcolor{Gray}
          \textbf{Electricity and Magnetism} & \textbf{Quantum Mechanics} & \textbf{Statistical Thermodynamics} \\ \hline 
         \textbf{DSECOP 250}: Automatic Differentiation& \textbf{DSECOP 235}: Schrodinger + NN \newline \textbf{DSECOP 236}: Schrodinger + GP & \textbf{DSECOP 320}: Monte Carlo + NN \\ \hline
        \rowcolor{Gray}
         \textbf{Advanced Lab} & \textbf{Particle, Nuclear, or Medical Physics} & \textbf{ Any Course} \\ \hline
         \textbf{DSECOP 120}: Exploratory Data Analysis \newline \textbf{DSECOP 240}: Symbolic Regression \newline \textbf{DSECOP 280}: Automated Video Analysis \newline \textbf{DSECOP 330}: SVD & \textbf{DSECOP 101}: Intro to DS Libraries \newline \textbf{DSECOP 110}: Intro to Histograms \newline\textbf{DSECOP 340}: NMR + NN & \textbf{DSECOP 102}: The Professor’s Module \newline \textbf{DSECOP 201}: Intro to Deep Learning \newline \textbf{DSECOP 210}: Intro to Random Forests \\ \hline
    \end{tabularx}
    \caption{An overview of courses in the undergraduate physics curriculum and existing modules for each course.}
    \label{tab:dsecop_courses}
\end{table}
\end{spacing}
}

\vspace{1em}

Our most introductory module is ``\ifthenelse{\boolean{anon_flag}}{COP}{DSECOP} 101: Introduction to Data Science Libraries" \ifthenelse{\boolean{anon_flag}}{(Please see Supplement\_1\_Intro\_to\_Data\_Science\_Libraries.zip)}{(\href{https://github.com/GDS-Education-Community-of-Practice/DSECOP/tree/main/Intro_to_Data_Science_Libraries}{link})}, which only assumes basic knowledge of Python and guides students through performing a simple data analysis with the Python libraries Pandas, Seaborn and Matplotlib. This module is connected to all other modules in Figure \ref{fig:dsecop_flowchart}. One of the more advanced modules offered, ``\ifthenelse{\boolean{anon_flag}}{COP}{DSECOP} 235: Schr\"{o}dinger's Equation and Neural Networks" (described in detail later in this paper), uses a complicated neural network to solve Schr\"{o}dinger's equation. If an instructor wanted to use this module in a course but feared their students' knowledge of neural networks was not strong enough, the instructor could also assign one or two of the prerequisite modules such as ``\ifthenelse{\boolean{anon_flag}}{COP}{DSECOP 230}: Differential Equations and Neural Networks" or ``\ifthenelse{\boolean{anon_flag}}{COP}{DSECOP} 201: Introduction to Deep Learning".

%% What modules currently exist and in what classes could they be used

The next two sections of this paper present two of these \ifthenelse{\boolean{anon_flag}}{COP}{DSECOP} modules in more detail. ``\ifthenelse{\boolean{anon_flag}}{COP}{DSECOP} 110: Introduction to Data Processing with Histograms" is one of our introductory modules giving an overview of creating meaningful graphs with histograms. ``\ifthenelse{\boolean{anon_flag}}{COP}{DSECOP} 235: Scr\"{o}dinger's Equation and Neural Networks" is an example of a more advanced module.

\subsubsection{Introduction to Data Processing with Histograms}

\ifthenelse{\boolean{anon_flag}}{Exhibit: Supplement\_2\_Intro\_to\_Data\_Processing\_with\_Histograms.zip}{Link: \url{https://github.com/GDS-Education-Community-of-Practice/DSECOP/tree/main/Intro_to_Data_Processing_with_Histograms}}

This module, designed for an undergraduate laboratory or a particle physics course, teaches students the steps involved in a data analysis pipeline. It focuses on key concepts such as processing datasets, creating histograms, curve fitting, and determining the goodness of fit for a chosen model and covers the practical use of tools such as NumPy and SciPy for creating histograms and curve fitting.

The module comprises in-class notebooks, a quiz, a homework assignment, and relevant datasets. Four notebooks guide students through the process of analyzing a dataset, including the creation of histograms and curve fitting. A short quiz assesses students' understanding. The homework assignment then requires students to undertake a practical data analysis task drawing from the material presented in the notebooks. Several datasets are provided for this purpose, along with Jupyter notebooks for instructors to re-generate the datasets if desired. It is estimated that the completion of each notebook will take approximately 60 minutes. The quiz is designed to take 30 minutes, and the homework assignment is anticipated to require less than 90 minutes of commitment.

The module explores the application of histograms as a tool for data visualization, where students are expected to create their own histogram from a raw dataset and employ built-in functions. Students are introduced to the concept of interpreting histograms as probability distributions. This concept is exemplified using a toy ``Uranium-241" dataset. Students are expected to import the data, create a histogram, and normalize these histograms.  

A bonus advanced topic available to students is the utilization of the Kolmogorov-Smirnov goodness-of-fit test for histograms, providing an additional level of depth for those wishing to further their understanding. Prerequisites for this module include familiarity with Python, numpy, and matplotlib.

\subsubsection{Learning the Schr\"odinger Equation}

\ifthenelse{\boolean{anon_flag}}{Exhibit: Supplement\_3\_Learning\_the\_Schrodinger\_Equation.zip}{Link: \url{https://github.com/GDS-Education-Community-of-Practice/DSECOP/tree/main/Learning_the_Schrodinger_Equation}}

This module aims to introduce students to deep learning as applied to the Time-Dependent Schr\"odinger Equation (TDSE) through a combination of analytical and machine learning methods. The module highlights physics-informed neural networks, a class of machine learning algorithms that facilitate the solution of partial differential equations by integrating constraints from partial differential equations within deep learning models. The module is designed for an undergraduate quantum mechanics course, and it guides students through solving the TDSE with a focus on the quantum harmonic oscillator. The only prerequisite is an introductory understanding of Python and the basics covered in the initial weeks of a quantum mechanics course.

The module is divided into three submodules. The first is a gentle, interactive introduction to deep learning, where students learn to create neural network pipelines through a simple example involving RGB color mixing. The second part offers an overview of various ML concepts with an emphasis on physics applications. The third submodule provides an interactive notebook for solving the TDSE for a 1D quantum harmonic oscillator system for the time evolution of a superposition of two energy eigenstates.

Supplementing the main content within the notebooks are hands-on lessons, exercises, homework, and suggested projects. These materials allow students to apply what they have learned and deepen their understanding of the subject. There are code cells that students are expected to fill themselves, references for further study, take-home assignments, and suggestions for extensions, such as the particle-in-a-box or the double-well potential, that build on the material and can be used as research projects. The estimated time for students to complete the module is about six hours, in addition to two hours taught by a professor in a classroom setting.

\section{Conclusions and Outlook}

In 2022, we introduced \ifthenelse{\boolean{anon_flag}}{COP}{DSECOP} as a means to address the need for the integration of data science into the undergraduate physics curriculum. Educators interested in including data science in their courses are welcome to join the community at \href{https://dsecop.org}{https://dsecop.org}. They can choose units based on their teaching subjects and their comfort with the methodology and are also welcome to engage with the community by contributing modules based on their use cases. We also hold workshops at conferences where instructors can interact more directly with the graduate fellows who develop our modules.

The DSECOP initiative represents a significant step toward modernizing physics education by embedding essential data science skills within the curriculum. This effort enhances the relevance of physics education and equips students with the tools necessary for a data-driven world. By creating a community of practice, we ensure that educators are supported and can share best practices, thereby continuously improving the integration of data science into physics education. The flexibility and accessibility of the modules developed by DSECOP are crucial for overcoming the challenges identified in our surveys, making data science education feasible even in resource-constrained environments. Through this initiative, we aim to create a robust foundation for future physicists who are well-versed in data science, thereby bridging the gap between traditional physics education and the demands of contemporary scientific research and industry.

% \begin{itemize}
% \item Emphasize the importance of data science in undergraduate physics education
% \item Provide actionable recommendations for physics teachers who want to incorporate data science into their courses
% \item Vision statement
% \end{itemize}

\ifthenelse{\boolean{anon_flag}}
{
% no Acknowledgements for double-anonymous review
}{
\section*{Acknowledgements}
The DSECOP project was funded and actualized with the support of the 2021 American Physical Society Innovation Fund award. This project would not have been possible without the support of Dr. Marilena Longobardi and Dr. Wolfgang Losert. We acknowledge useful discussions with Dr. Jackson and Dr. Craig from EP3, Dr. Robert Hilborn from AAPT, and Dr. Talitha Washington from Atlanta University Center. We extend our acknowledgment to our DSECOP Fellows for their invaluable contributions, without which this project would not have achieved its success; 2022 fellows, listed in alphabetical order: Sebastian Atalla, Fatemeh Bagheri, Julie Butler, Cunwei Fan, Radha Mastandrea, and Karan Shah; and 2023 fellows, also in alphabetical order: Julie Butler, Ashley Dale, Joseph Dominicus Lap, Richard Harry, Connor Robertson, and Karan Shah. Support for Dr. William Ratcliff’s participation in this education and outreach activity was provided by the CHRNS project under a partnership between the NIST and the NSF (DMR-2010792). Any mention of commercial products within this paper is for information only; it does not imply recommendation or endorsement by NIST.
}

\section*{Conflict of Interest}

The authors have no conflicts to disclose.

\bibliographystyle{unsrt}
\bibliography{ajp_dsecop}

\section*{Appendix: Survey Details}

In this section, we describe additional faculty survey details. 8\% of the faculty indicated they teach data science in their introductory and intermediate/advanced physics courses. Throughout this manuscript, both intermediate and advanced physics courses will be referred to as ``advanced" for brevity. 2\% indicated they teach data science only in introductory physics. 5\% indicated that their introductory physics course that teaches data science is focused on data science (e.g., Data Science for Physicists). 19\% indicated they teach data science in only intermediate or advanced physics. 12\% indicated that their advanced physics course is focused on data science.

For introductory courses, data science is taught via in-class activities (\textit{N} = 20), projects (\textit{N} = 8), and homework exercises (\textit{N} = 7). A few respondents include data science on quizzes or exams (\textit{N} = 3). 

For advanced courses, data science is taught via in-class activities (\textit{N} = 8), projects (\textit{N} = 25), and homework exercises (\textit{N} = 19). Some respondents include data science questions on quizzes or exams (\textit{N} = 8). 

The majority of respondents (\textit{N} = 70) have not taught data science in their physics courses in the past 5 years. Out of the 70 respondents, 69 answered the question regarding their interest in teaching data science in their physics courses. 40.6\% selected they would be interested, 40.6\% were maybe, and 18.8\% were not interested. Sixty-one of the respondents answered the free-response question regarding why they do not teach data science in their physics courses. These responses were thematically coded. The respondents often gave multiple reasons. The top themes that emerged are in the table \ref{tab:FacultyResponses} below.

\begin{spacing}{0.8}
\begin{table}
    \centering
    \begin{tabularx}{\linewidth}{|>{\hsize=0.8\hsize}X|>{\hsize=0.4\hsize}X|>{\hsize=1.8\hsize}X|} 
    \hline
        \rowcolor{Gray}
         \textbf{Theme} & \textbf{Count} & \textbf{Example quotes}  
         
         \\ \hline
         {Not relevant or necessary courses taught} & {20} & ``... I have not judged it to be a priority (or even relevant) in courses I am teaching on topics such as quantum mechanics, special relativity, and `honors' level mechanics.''\newline \newline {``Students need to learn the conceptual framework of physics first.  There is little point in teaching them to compute if they have nothing to compute.''} \newline \\ \hline

        {Challenges in changing or adding to an already full curriculum}  & {18}& {``Courses are already full. Hard to decide what to drop to make room'' \newline \newline} {``No space in the curriculum.  Need to focus on teaching students physics concepts and methods, as well as some basic computational skills.''} \newline \newline {``...not sure how to incorporate since we don't have room in the program for a separate course ''} \newline \\ \hline

         {Faculty do not know how to use data science} & {13} & ``I don't understand machine learning well enough to teach it. Although, it seems no one does.'' \newline \newline {``No background in big data, ML, or AI. ''} \newline \\ \hline

          {Faculty does not know how to include data in their physics courses} & {13} & ``I don't know how I would integrate it into courses I have taught...''\newline \newline {``I have no direct experience, wouldn't know where to start or what courses to add it to.''} \newline \\ \hline
    \end{tabularx}
    \caption{Thematic coding of faculty response regarding why they do not include data science in their undergraduate physics courses}
    \label{tab:FacultyResponses}
\end{table}
\end{spacing}

Most respondents did not see data science as relevant or necessary to the courses they teach. These respondents were not opposed to data science; some used data science and AI in their research, but they felt that teaching such skills was part of the core curriculum or could be learned elsewhere (e.g., in a computer science department). This is related to the second most identified theme, that there are challenges in changing or adding to the physics curriculum. These respondents believed that the curriculum has a lot of material to cover as is, and some were already engaging in changes. Lastly, some respondents do not have experience using data science, and others were unsure how to include it in their courses.

Some other themes include students having little experience programming (\textit{N} = 9), faculty bandwidth to change the curriculum in their courses (\textit{N} = 4), and lack of decision-making power to change the curriculum (\textit{N} = 3).

\end{document}